\documentclass[aps,reprint,superscriptaddress]{revtex4-1}

\usepackage{graphicx,color}
\usepackage{graphics}
\usepackage{lipsum}
\usepackage{float}
\floatstyle{boxed}
\usepackage{blkarray}
\usepackage{flushend}

\begin{document}


\title{Impact of g-factors and valleys on spin qubits in a silicon double quantum dot}


\author{J. C. C. Hwang}
\email{jason.hwang@unsw.edu.au}
\author{C. H. Yang}
\affiliation{Centre for Quantum Computation and Communication Technology, School of Electrical Engineering and Telecommunications, The University of New South Wales, Sydney NSW 2052, Australia}

\author{M. Veldhorst}
\affiliation{Centre for Quantum Computation and Communication Technology, School of Electrical Engineering and Telecommunications, The University of New South Wales, Sydney NSW 2052, Australia}
\affiliation{QuTech, TU Delft, 2600 GA Delft, The Netherlands}

\author{N. Hendrickx}
\affiliation{University of Twente, PO Box 217, 7500 AE Enschede, The Netherlands}

\author{M. A. Fogarty}
\author{W. Huang}
\affiliation{Centre for Quantum Computation and Communication Technology, School of Electrical Engineering and Telecommunications, The University of New South Wales, Sydney NSW 2052, Australia}
\author{F. E. Hudson}
\author{A. Morello}
\author{A. S. Dzurak}
\email{a.dzurak@unsw.edu.au}
\affiliation{Centre for Quantum Computation and Communication Technology, School of Electrical Engineering and Telecommunications, The University of New South Wales, Sydney NSW 2052, Australia}


\date{\today}

\begin{abstract}
We define single electron spin qubits in a silicon MOS double quantum dot system. By mapping the qubit resonance frequency as a function of gate-induced electric field, the spectrum reveals an anticrossing that is consistent with an inter-valley spin-orbit coupling. We fit the data from which we extract an inter-valley coupling strength of 43 MHz. In addition, we observe a narrow resonance near the primary qubit resonance when we operate the device in the (1,1) charge configuration. The experimental data is consistent with a simulation involving two weakly exchanged-coupled spins with a Zeeman energy difference of 1 MHz, of the same order as the Rabi frequency. We conclude that the narrow resonance is the result of driven transitions between the $T_{-}$ and $T_{+}$ triplet states, using an ESR signal of frequency located halfway between the resonance frequencies of the two individual spins. The findings presented here offer an alternative  method of implementing two-qubit gates, of relevance to the operation of larger scale spin qubit systems. 
\end{abstract}


\maketitle


The seminal proposal by Loss and DiVincenzo~\cite{Loss1998} for spin-based quantum computing using semiconductor quantum dots has led to numerous experimental demonstrations~\cite{Petta2005,Koppens2006,Nowack2011,Shulman2012,Medford2013} and helped inspire the growing field of quantum spintronics~\cite{Awschalom2013}. Progress in silicon quantum dot qubits~\cite{Maune2012,Kim2014,Kawakami2014} has established promising coherence times, with as long as 28 ms~\cite{Veldhorst2014} being achieved in  isotopically purified $^{28}$Si substrates~\cite{Itoh2014}. The use of silicon as a device platform also has the advantage of sharing many similarities with  standard manufacturing technologies used in today’s microelectronics industry. Recently, universal quantum logic~\cite{Nielsen2000} in silicon has been demonstrated  via the realization of single-qubit~\cite{Veldhorst2014} and two-qubit logic gates~\cite{Veldhorst2015}, opening the path towards multi-qubit coherent operations in silicon.

Previous work~\cite{Veldhorst2015} illustrates that individual silicon quantum dots can possess local variability in g-factor. By exploiting a gate-induced Stark shift, the g-factors of neighboring qubits can be tuned far apart with respect to their exchange interaction to enable high-fidelity CZ operations. In this letter, we analyze single electron spin qubits defined in a silicon metal-oxide-semiconductor (SiMOS) double quantum dot system and show that additional two-qubit gate operations can occur when the g-factors of neighboring qubits are close. In particular, our experimental data indicates that an electron spin resonance (ESR) frequency that is not in direct resonance with any of the individual qubits can simultaneously excite a pair of neighboring qubits. This effect has important implications for the scalability of silicon quantum dot systems, since while small g-factor differences may not provide the best setting for CZ operations~\cite{Veldhorst2015}, one could find it desirable and more flexible to operate two qubit gates using a single ESR signal.

\begin{figure*}
	\includegraphics[width=0.9\linewidth]{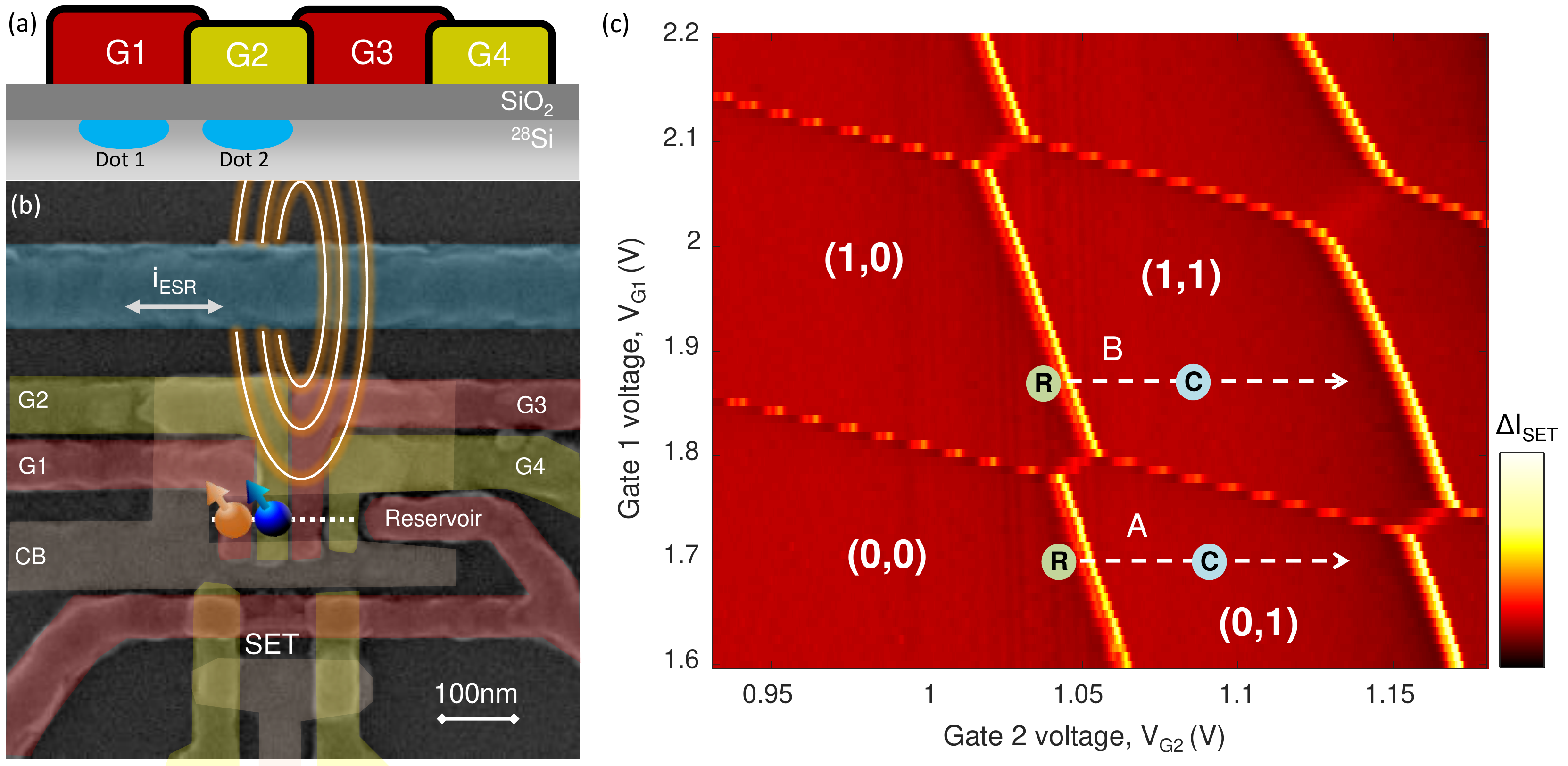}
	\caption{(a) Cross-sectional schematic along the white dotted line in (b), and (b) scanning electron microscope image of the SiMOS quantum dot device. Multi-layer aluminium gate electrodes are patterned using electron beam lithography. The aluminium oxide that serves as insulator between gates also forms a natural tunnel barrier between the double dots that  are formed under G1 and G2. An adjacent SET is used to monitor the charge occupancy of the dot while a microwave antenna allows coherent control of the qubit via ESR. (c) Charge stability diagram of a coupled double dot system, as a function of gate voltages $V_{G1}$ and $V_{G2}$. The charge occupancy of the relevant regions are labeled as (N1,N2), where N1 and N2 are the electron occupancies of dot 1 and dot 2 respectively.}
	\label{FIG1}
\end{figure*}

The quantum dot device, as shown in Fig.~\ref{FIG1}b, consists of aluminium gates fabricated on an isotopically purified silicon epilayer substrate via multi-layer gate stack technology~\cite{Angus2007,Veldhorst2014}. A single electron transistor (SET) for charge sensing is fabricated next to four control gates (G1 - G4), each of which can be independently tuned to locally define a quantum dot, as the aluminium oxide between  neighboring gates forms natural  tunnel barriers. Electrons are supplied to the quantum dots via a reservoir that is induced by a gate next to G4 that branches out from the SET top-gate, so that the reservoir is connected to the SET drain. An on-chip broadband microwave line~\cite{Dehollain2012} for generating a high-frequency oscillating magnetic field is fabricated parallel to the device.

Using electrostatic confinement, we create a double quantum dot system under G1 and G2. This is tunnel coupled to an electron reservoir which extends under G3 and G4, since both these gates are biased well above threshold. The electron occupancy of each dot is electrically controlled via voltages applied to the gates. Figure~\ref{FIG1}c shows a charge stability diagram of the system, measured by the nearby SET charge sensor~\cite{Yang2011}. We can deplete both dots, under G1 and G2 respectively, to their last electron. The charge transitions as a function of the voltages on gates G1 and G2 form a characteristic honey-comb pattern, which demonstrates the electrostatic coupling between the dots.

Here, the spin states of a single electron under a static external magnetic field $B_0$ = 1.45 T are separated by the Zeeman splitting, $E_{Z} = g\mu_{B} B_0$, where $g$ is the electron g-factor and $\mu_{B}$ is the Bohr magneton. By applying a high-frequency oscillating current through the on-chip microwave antenna, an oscillating magnetic field is generated. Coherent control of the qubit is achieved when the frequency of this a.c. magnetic field matches the electron Zeeman splitting. The spin state of the electron is measured in single shot via spin-to-charge conversion ~\cite{Elzerman2004}.

Two-qubit gates~\cite{Veldhorst2015} can be realized by initializing the qubits in the (1,1) charge state and switching on the exchange interaction via fast gate pulsing towards the (0,2) transition. In the current experiment, however, we only have high frequency pulse control of G2 and not G1. As a result, we cannot pulse diagonally towards the (1,1)--(0,2) transition. The high tunnel rate between dot 1 (under G1) and the reservoir also prohibits us from reading the spin state of this dot. Therefore in this work, we rely on pulsing and reading dot 2 only, to characterize the device in the (1,1) charge region.

\begin{figure}
	\includegraphics[width=\columnwidth]{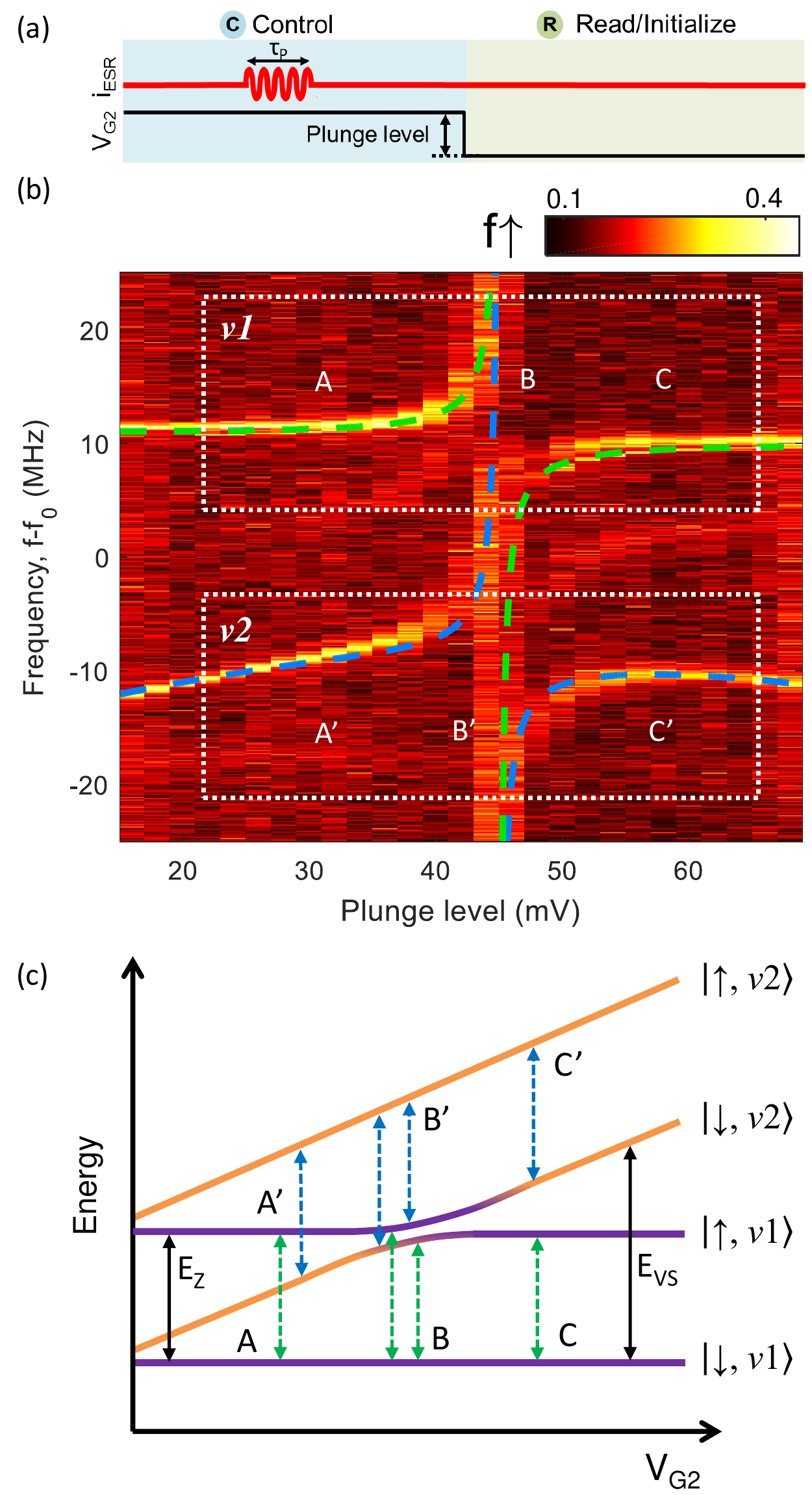}
	\caption{(a) Gate pulse sequence for ESR control. An ESR microwave burst of pulse length $\tau_{p}$ is applied to the qubit, followed by spin readout and initialisation of a spin-down electron for the next control pulse. (b) ESR spectrum, showing electron spin-up fraction f$\uparrow$ as a function of qubit resonance frequency and G2 plunge level. The external magnetic field is set at 1.45 T, with voltage operating point $V_{G1}$ = 1.87 V and $V_{G2}$ = 1.047 V. The resonance frequency $f_0$ = 40.23 GHz and $\tau_p = 50~\mu s$. The resonances are fitted by dashed lines, with a color code described by (c), which is a model of a quantum dot incorporating both spin and valley degrees of freedom.}	
	\label{FIG2}
\end{figure}

A standard two-level pulse sequence (see Fig.~\ref{FIG2}a) is applied to gate G2 in order to map out the resonance frequency of the electron spin qubit formed in dot 2 as a function of plunge level $V_{G2}$. Stark shifting of the electron g-factor via electric field has been experimentally reported in a similar device~\cite{Veldhorst2014,Veldhorst2015prb}, where the plunger gate voltage has direct control over the out-of-plane electric field through the quantum dot, and hence the qubit resonance frequency. In this device, an expected linear dependency of qubit resonance frequency on the plunge level is observed (Fig.~\ref{Stark} in Appendix~\ref{AppA}) when the qubit is operating in the (0,1) charge region, as marked by A in Fig.~\ref{FIG1}c. Here (C) and (R) represent the qubit control and readout position respectively.

When the qubit system is operated in the (1,1) charge region, labelled B in Fig.~\ref{FIG1}c, we obtain an ESR spectrum (Fig.~\ref{FIG2}b) that contains multiple resonance branches. Coherent Rabi oscillations can be obtained at the two bright branches. Anti-crossings in the frequency spectrum are also revealed, which are the result of coupling to another degree of freedom. To investigate the origin of these anti-crossings, the ESR spectrum was mapped out at several different values of $V_{G1}$ and at two different magnetic fields, and the corresponding location of the anti-crossing point was measured in terms of the plunge level $V_{G2}$ where ESR was performed. Fig.~\ref{AntiCrossing} in Appendix~\ref{AppB} shows the anti-crossing location on the charge stability diagram.

Due to the gate voltage dependence of the anti-crossing, the additional state is likely to be another charge state, or an excited valley state, and we consider the likelihood of each possibility in turn. We first consider the possibility of a charge transition. It is immediately clear (from Fig.~\ref{AntiCrossing} in Appendix~\ref{AppB}) that the locus of the anti-crossing occupies the center of the (1,1) charge region, indicating that a charge state is unlikely to be the cause, as this would require the ESR plunge level to be very close to the (1,1)--(0,2) or (1,1)--(1,2) charge transition. This is further rejected by the observation of the resonances bending upward to the left of the anti-crossing and downward to the right, which indicates the energy of the state increases with increasing $V_{G2}$ gate voltage, a trend that is directly opposite to what a (0,2) or (1,2) charge state would exhibit when $\left|\downarrow,\downarrow\right\rangle$ is the ground state of the system~\cite{Veldhorst2015}.

Next we consider the possibility of an excited valley state, and note that a similar type of anti-crossing has been observed previously~\cite{Jiang2014} and can be attributed to inter-valley spin-orbit coupling~\cite{Jiang2014}, which occurs when the valley splitting equals the Zeeman splitting, $E_{VS} = E_Z$. The six-fold valley degeneracy in the conduction band of bulk silicon is lifted via confinement of electrons in a quantum dot, leaving two low-lying valley states with energy scales relevant to spin qubit operation. It has been experimentally demonstrated that the valley splitting $E_{VS}$ in a quantum dot is dependent on the out-of-plane electric field~\cite{Yang2013} and can be controlled using gate potential over a range of 0.5 meV~\cite{Yang2013,Veldhorst2014}. 

Figure~\ref{FIG2}c shows the energy level diagram of a model based on valley states we devised for generating the fittings (blue and green dotted lines) overlaid on the experimental data in Fig.~\ref{FIG2}b. The fit Hamiltonian is included in Appendix~\ref{AppC}. The model assumes a single quantum dot with $E_{VS}$ tunable via the voltage on G2. Mixing between $\left|\uparrow, v1\right\rangle$ and $\left|\downarrow, v2\right\rangle$ states occurs when the valley splitting energy approximately equals the Zeeman energy, and modifies the resonance frequency of the qubit. From the model we extract an inter-valley coupling strength $\beta$ =
43 MHz. We find experimentally that a 40 mV change on $V_{G2}$ is required to offset the anti-crossing energy by an equivalent amount to a change in magnetic field of 0.1 T (see Fig. S2). In a similar SiMOS quantum dot device~\cite{Yang2013} the same level of energy tuning required a 18 mV change on the plunger gate potential. The energy tuning for both devices is of a similar order of magnitude, and the small difference is most likely attributed to differences in the voltage biasing arrangement between the two devices.

An ESR driven spin transition within a valley should produce only a single resonance in the spectrum; our observation of two resonances could be explained if we assume that the qubit can be initialized to the spin-down state of either valley~\cite{Kawakami2014}, and that the two valleys have a g-factor difference of approximately 20 MHz. During spin readout/initialization, $E_{VS}$ is smaller than $E_{Z}$, and so the Fermi level of the reservoir can be positioned between the spin-down and -up states of the two valleys (left side of Fig. 2c). $E_{VS}$ is then subsequently increased due to the deeper plunge level during ESR control, where spin qubit is driven in either one of the two  valleys. With the aforementioned tunability of $E_{VS}/g\mu_B$ corresponding to 0.1 T per 40 mV on the plunger gate, this implies that at the readout position, the gap for differentiating between  $\left|\downarrow, v2\right\rangle$ and $\left|\uparrow, v1\right\rangle$ is around $g\mu_B$(0.11 T) $\approx 13~\mu eV$. The valley initialization assumption further indicates the valley relaxation time in our system is long, at least longer than the dot 2 plunge time of 950 $\mu$s. This falls within the range of possible valley relaxation times predicted recently in Ref.~\cite{Boross2016}, where the valley relaxation rate is estimated to be a strong function of the relative location of the quantum dot to a step at the Si-SiO$_2$ interface.

\begin{figure}
	\includegraphics[width=\columnwidth]{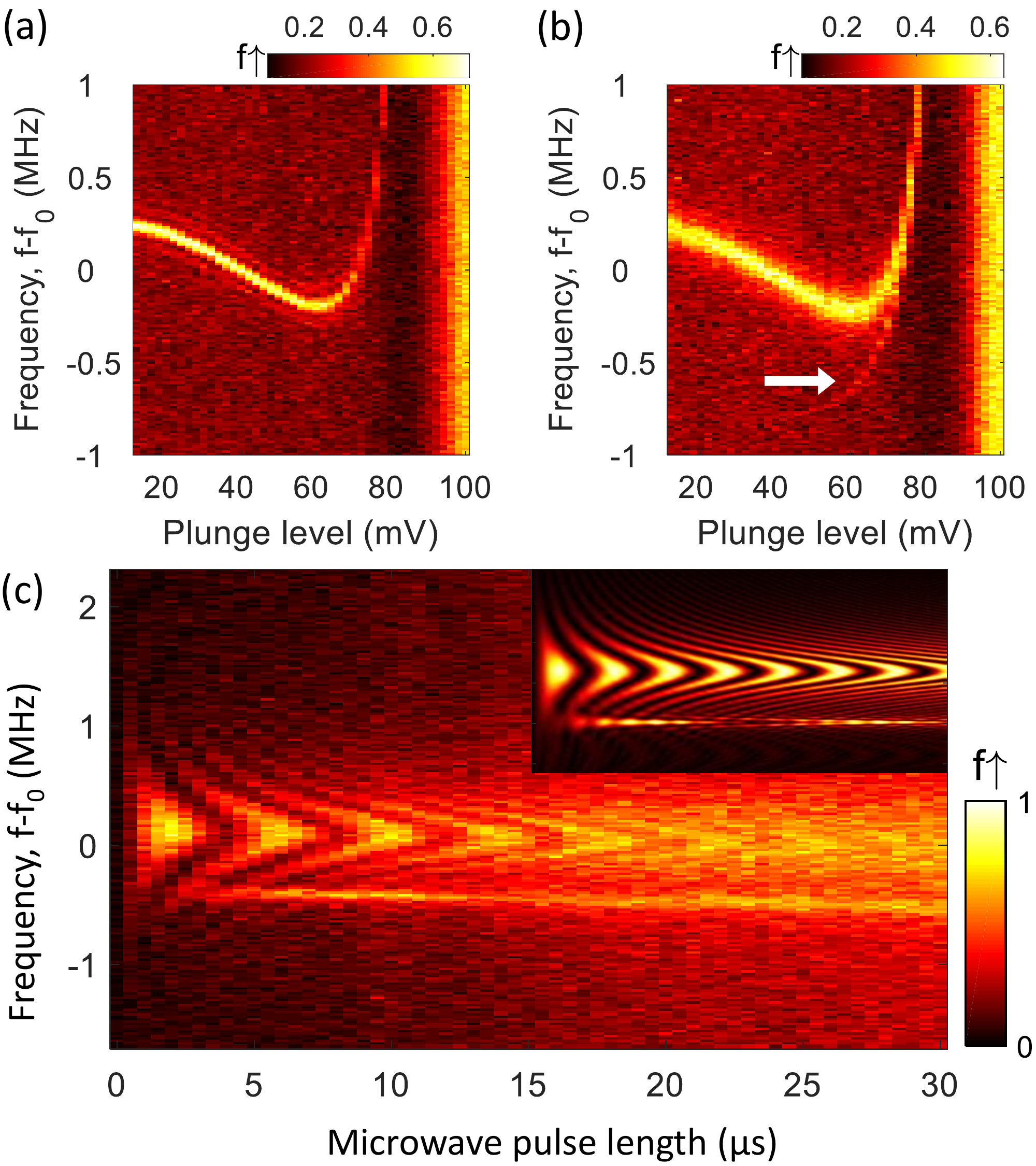}	
	\caption{(a) Electron spin-up fraction f$\uparrow$ as a function of qubit resonance frequency and ESR plunge level, with $V_{G1}$ = 1.87 V and $f_0$ = 42.947 GHz. The ESR microwave pulse length is applied for 14 $\mu$s. Clear non-linearity in the resonance branch can be seen. (b) An additional level appears beneath the main branch when the same measurement is repeated with a longer microwave pulse length of 214 $\mu$s. (c) Rabi oscillation obtained at $V_{G1}$ = 1.82 V, with $f_0$ = 42.972 GHz. Adjacent to the main Rabi chevron pattern is a narrow oscillation branch, which is associated with the faint resonance observed in (b). Inset: the corresponding simulated Rabi oscillation by measuring only dot 2, with Zeeman energy difference between the two dots $\delta E_Z$ = 1 MHz and Heisenberg exchange coupling J = 250 kHz.}	
	
	\label{FIG3}
\end{figure}

We now focus our attention on the small frequency splitting that is seen in the upper resonance branch before and after the anti-crossing. We zoom into that region and perform higher resolution mapping of the ESR spectrum, as shown in Fig.~\ref{FIG3}a, with the magnetic field set at 1.55 T and the microwave pulse length at 14 $\mu$s. The anti-crossing seen in Fig.~\ref{FIG2}b is shifted in location as the gate voltages of G1 and G2 are changed. The dramatic increase in the spin-up fraction at the far right of the map is simply the result of changes in the SET current level due to loading of another electron as we approach the (1,1)--(1,2) charge transition using a deeper plunge level.

Repeating the same measurement with longer microwave pulse length reveals an additional resonance which appears near the bending of the resonance branch, as shown in Fig.~\ref{FIG3}b. This divergent resonance corresponds to the narrow resonance next to the primary Rabi chevron pattern in Fig.~\ref{FIG3}c, where we plot the electron spin-up fraction as a function of microwave pulse length and frequency detuning. 

As we do not observe this extra resonance frequency when the qubit is operated in the (0,1) charge region, this lead us to the belief that we observe an effect related to coupling with the adjacent qubit in dot 1. Indeed, we can closely match the experimental data with a simulated Rabi oscillation (inset of Fig.~\ref{FIG3}c) in which we assume that the spin in dot 2 is weakly exchange coupled to that in dot 1 (J = 250 kHz) where the two dots have a different Zeeman energy $\delta E_Z = \delta g\mu_B B_0$ = 1 MHz due to their difference in g-factor, $\delta g$. The system Hamiltonian in the rotating-wave approximation is given in Eq. 1, with basis $\{\left|\uparrow,\uparrow\right\rangle$, $\left|\uparrow,\downarrow\right\rangle$, $\left|\downarrow,\uparrow\right\rangle$, $\left|\downarrow,\downarrow\right\rangle\}$:

\begin{equation}
\left(\begin{array}{cccc}
-\Delta \omega - \frac{1}{2}\delta E_Z & \Omega & \Omega & 0 \\
\Omega & \frac{1}{2}\delta E_Z -\frac{1}{2}J & \frac{1}{2} J & \Omega  \\
\Omega & \frac{1}{2} J & -\frac{1}{2}\delta E_Z -\frac{1}{2}J & \Omega\\
0 & \Omega & \Omega & \Delta \omega + \frac{1}{2}\delta E_Z\\
\end{array}\right)
\label{Eqn:Rabi}
\end{equation}  
where $\Delta \omega$ is the microwave frequency minus the Larmor frequency of spin 2, $\Omega$ is the Rabi frequency. $J = \frac{2t^2}{U-\epsilon-\delta E_Z}+\frac{2t^2}{U-\epsilon+\delta E_Z}$, as derived in the Supplementary Information of Ref~\cite{Veldhorst2015}, is an effective exchange arising from tunnel coupling between the $\left|\uparrow,\downarrow\right\rangle$, $\left|\downarrow,\uparrow\right\rangle$ state and the (0,2) state, where U is the on-site Coulomb energy and $\epsilon$ the detuning. The simulation only measures the z-component of spin 2, which corresponds to reading out only dot 2 in the experiment. Further details on the Rabi simulation are included in Appendix~\ref{AppD} and~\ref{AppE}.
 
When the applied microwave frequency is halfway between the two qubits' resonance frequencies, and given that there is a finite exchange coupling, the two qubits are excited simultaneously from $T_{-}$ ($\left|\downarrow,\downarrow\right\rangle$) to $T_{+}$($\left|\uparrow,\uparrow\right\rangle$). Since we only perform readout on dot 2, the ESR chevron pattern of the other spin is absent from Fig.~\ref{FIG3}c, and the narrow-band resonance corresponds to the flipping of $T_{-}$ to $T_{+}$. The effective coupling $C_{\rm eff}$ between the two triplet states originates from a second order effect via the $\left|\uparrow,\downarrow\right\rangle$ and $\left|\downarrow,\uparrow\right\rangle$ states. The resulting Hamiltonian approximated from second-order perturbation on Eq.1 is given as follows ($\delta E_Z > J$ for the approximation to hold):  
\begin{equation}
\left(\begin{array}{cc}
-\Delta \omega - \frac{1}{2}\delta E_Z & C_{\rm eff} \\
C_{\rm eff} & \Delta \omega + \frac{1}{2}\delta E_Z
\end{array}\right)
\left(\begin{array}{c}
\left|\uparrow,\uparrow\right\rangle \\
\left|\downarrow,\downarrow\right\rangle
\end{array}\right)
\end{equation}
where $C_{\rm eff}$ $= 4 J \Omega^2/(\delta E_Z^2 - J^2)$ (see Appendix~\ref{AppE} for derivation). This indicates the coupling diminishes to zero for large $\delta E_Z$ and only becomes visible when $\delta E_Z$ reduces to a similar order of magnitude as the exchange. 

\begin{figure}[h]
	\includegraphics[width = 1\columnwidth]{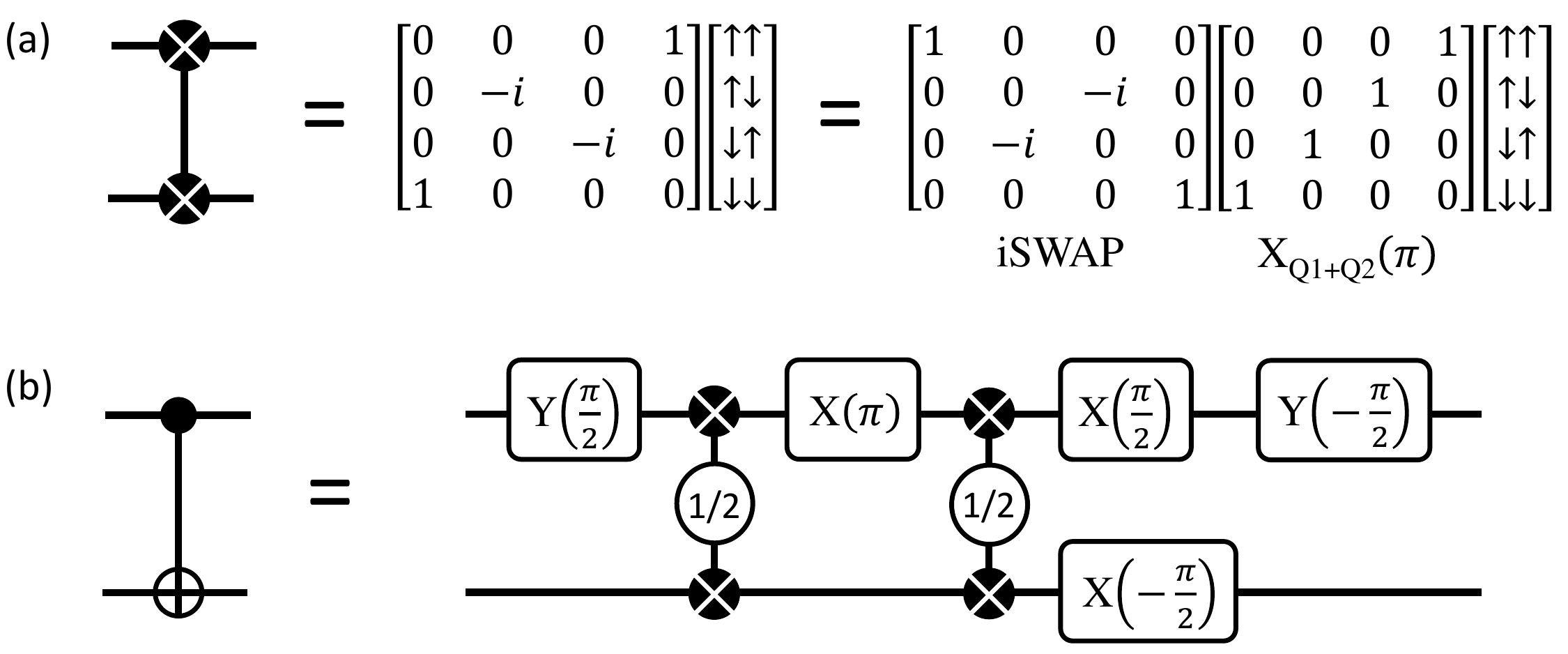}%
	\caption{(a) The two-spin rotation is a combination of $\rm iSWAP$ and X($\pi$) on both qubits. (b) The decomposition of a CNOT gate into two-spin rotations and single qubit gates. The 1/2 denotes the application of the two-qubit gate in (a) with half the duration.}	
	\label{iSWAP}
\end{figure}

The two-spin rotation here is a combination of $\rm {iSWAP}$ and X($\rm \pi$) rotation on both qubits. By applying half the duration of the microwave pulse, a universal two-qubit gate can be realized. For example a CNOT can be constructed as shown in Fig.~\ref{iSWAP}. 

In conclusion, we have analyzed a spin qubit formed in a SiMOS double quantum dot system, where the qubit is weakly exchange coupled to a neighboring spin. The ESR spectrum shows an anti-crossing in the resonance frequency that is consistent with an inter-valley spin-orbit coupling, with a strength of 43 MHz. Previous qubit devices~\cite{Veldhorst2015} reported a $\delta E_Z$ between two neighboring dots that varies between 20 to 40 MHz. The findings here also reveal a new mechanism that can be exploited for qubit operations when the g-factor difference is small, with our simulation result suggesting a $\delta E_Z$ of 1 MHz. This has allowed us to observe ESR-driven transitions between the $T_{-}$ and $T_{+}$ state, which requires only the use of a single ESR pulse to simultaneously rotate two individual spins. This could be used in future to add flexibility to qubit operations in large-scale silicon quantum dot systems.

\begin{acknowledgments}
The authors thank D. Culcer for enlightening discussions. We thank D. Barber and R. Ormeno for technical support in the measurement laboratory. This work was supported by the Australian Research Council (CE11E0001017), the U.S. Army Research Office (W911NF-13-1-0024), and the NSW Node of Australian National Fabrication Facility.
\end{acknowledgments}

\appendix
\section{Linear Stark Shift of Qubit Resonance Frequency}
\label{AppA}
\begin{figure}[h]
	\includegraphics[width=0.4\textwidth]{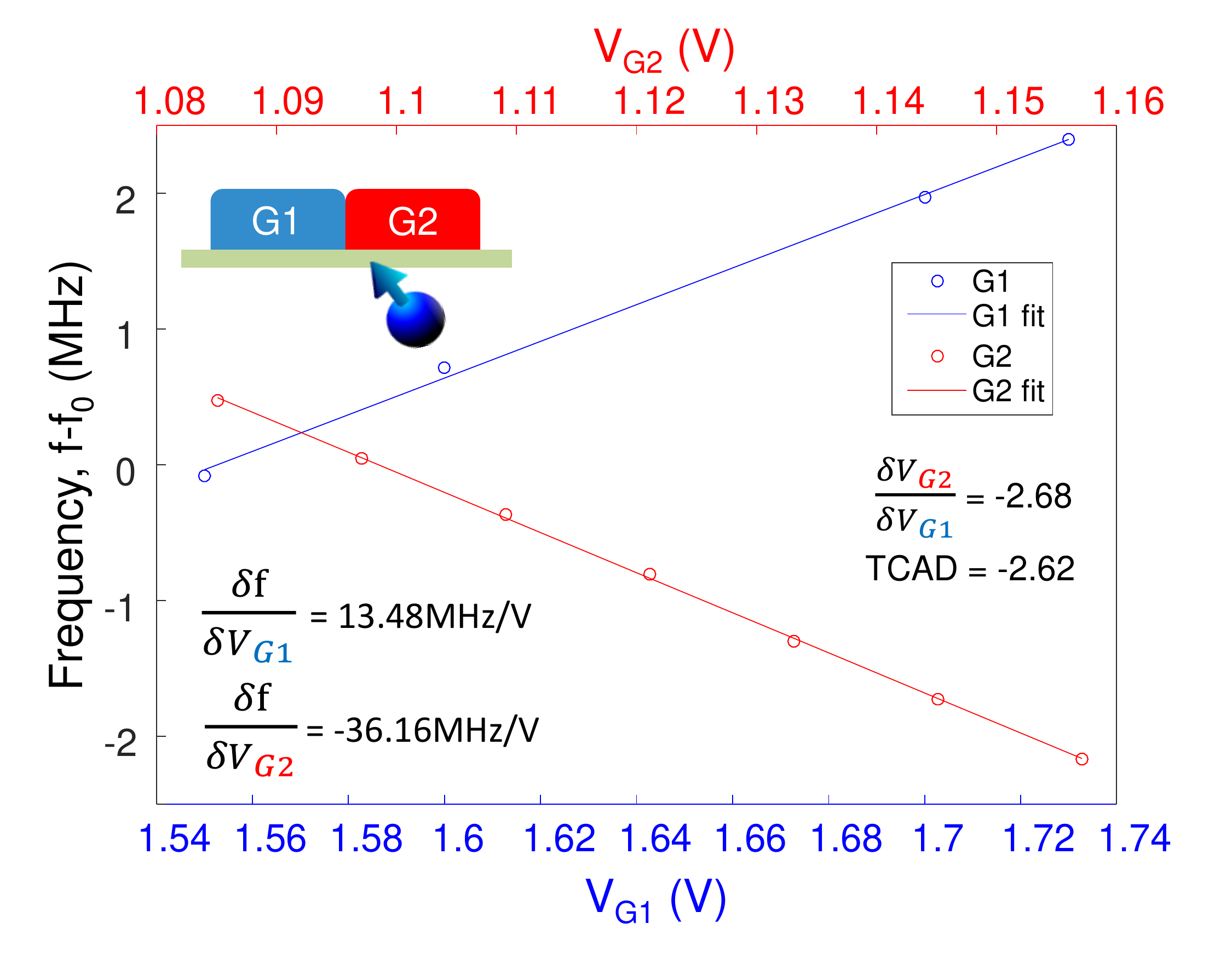}
	\caption{Qubit resonance frequency as a function of barrier gate voltage G1 (blue) and plunger gate G2 (red).} 
	\label{Stark}
\end{figure}

When the qubit is operated in the (0,1) charge region as marked by trajectory (A) in Fig. 1b, the resonance frequency as a function of both the plunger gate and the barrier gate voltages are found to be linear, with their relative contribution to the electric field $\frac{dF/dV_{G1}} {dF/dV_{G2}}$ = $R_{F,V_{G1}V_{G2}}$ = -2.68 (See Fig S1). We have performed simulations using the Synopsys Sentaurus Semiconductor TCAD Software and found a ratio $R_{F,V_{G1}V_{G2}}$ = -2.62, which shows excellent agreement with the experimental value.

\section{Anticrossing Location}
\label{AppB}
\begin{figure}[h]
	\includegraphics[width = 1\columnwidth]{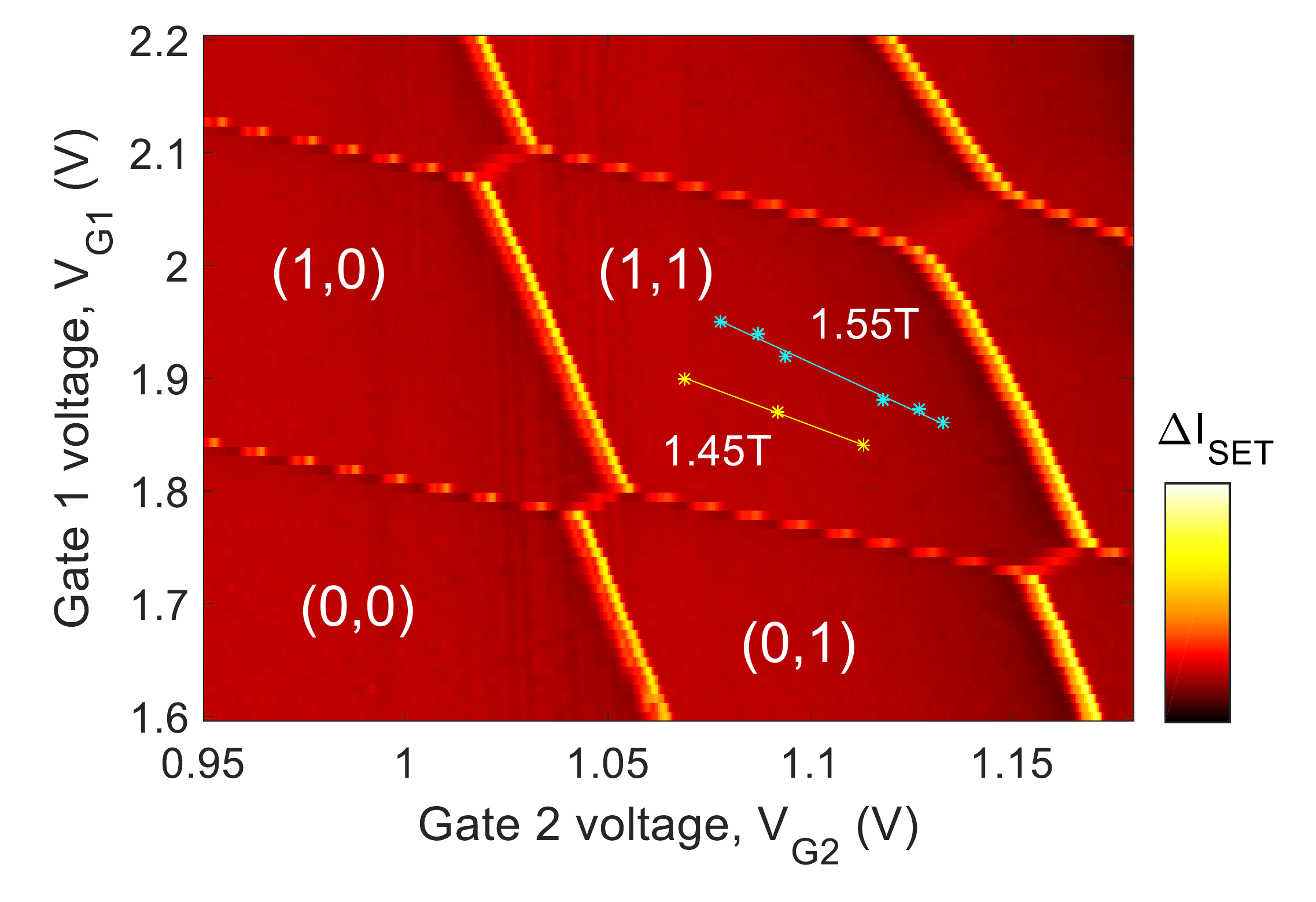}%
	\caption{Anti-crossing location as a function of $V_{G1}$. The locus is plot out for two different magnetic field magnitude.}	
	\label{AntiCrossing}
\end{figure}

The location of the anticrossing in Fig.~\ref{FIG2} in the main text is highly dependent on gate voltages. We perform similar ESR spectrum mapping at several different gate voltages of G1 and measure the voltage G2 required to plunge from the spin readout level to reach the anticrossing. The experiment is also performed under two different external magnetic field (1.55T and 1.45T) and the resulting anticrossing position is plotted on the charge stability map in Fig.~\ref{AntiCrossing}. 

\section{Fitting of the ESR Spectrum}
\label{AppC}
The fitting of Fig. 2 in the main text is achieved with a model that considers the spin states of a single electron, Zeeman splitted by a DC magnetic field, and takes into account of the valley degree of freedom. The static Hamiltonian is a simple $4 \times 4$ matrix, with the basis being $\left|\uparrow,v1\right\rangle$, $\left|\uparrow,v2\right\rangle$, $\left|\downarrow,v1\right\rangle$, $\left|\downarrow,v2\right\rangle$: 

\begin{equation}
H = 
\left(\begin{array}{cccc}
\frac{E_{Z,v1}+E_{Z,v2}}{2}+E_{VS} & 0 & 0 & 0 \\
0 & \frac{E_{Z,v1}-E_{Z,v2}}{2}+E_{VS} & \beta & 0 \\
0 & \beta & E_{Z,v2} & 0\\
0 & 0 & 0 & 0\\
\end{array}\right)
\label{Eqn:FitModel}	
\end{equation}

where $E_{Z,v1}$ and $E_{Z,v2}$ are the Zeeman splitting in the upper and lower valley. The valley splitting $E_{VS}$ is dependent on the electric field through the dot, which is in turn dependent on the gate voltage. In the model we assume a valley splitting tunability of 640 $\mu$eV/V as quoted in ref [21]. Valley states become relevant when either $E_{VS}$ is very small or when $E_{VS}$ is in the order of the Zeeman splitting, with the latter being the case for our fitting. $\beta$ is the inter-valley coupling parameter, which involves a spin-flip.\\ 

\section{Rabi Simulation}
\label{AppD}
\begin{figure}[h]
	\includegraphics[width = 0.8\columnwidth]{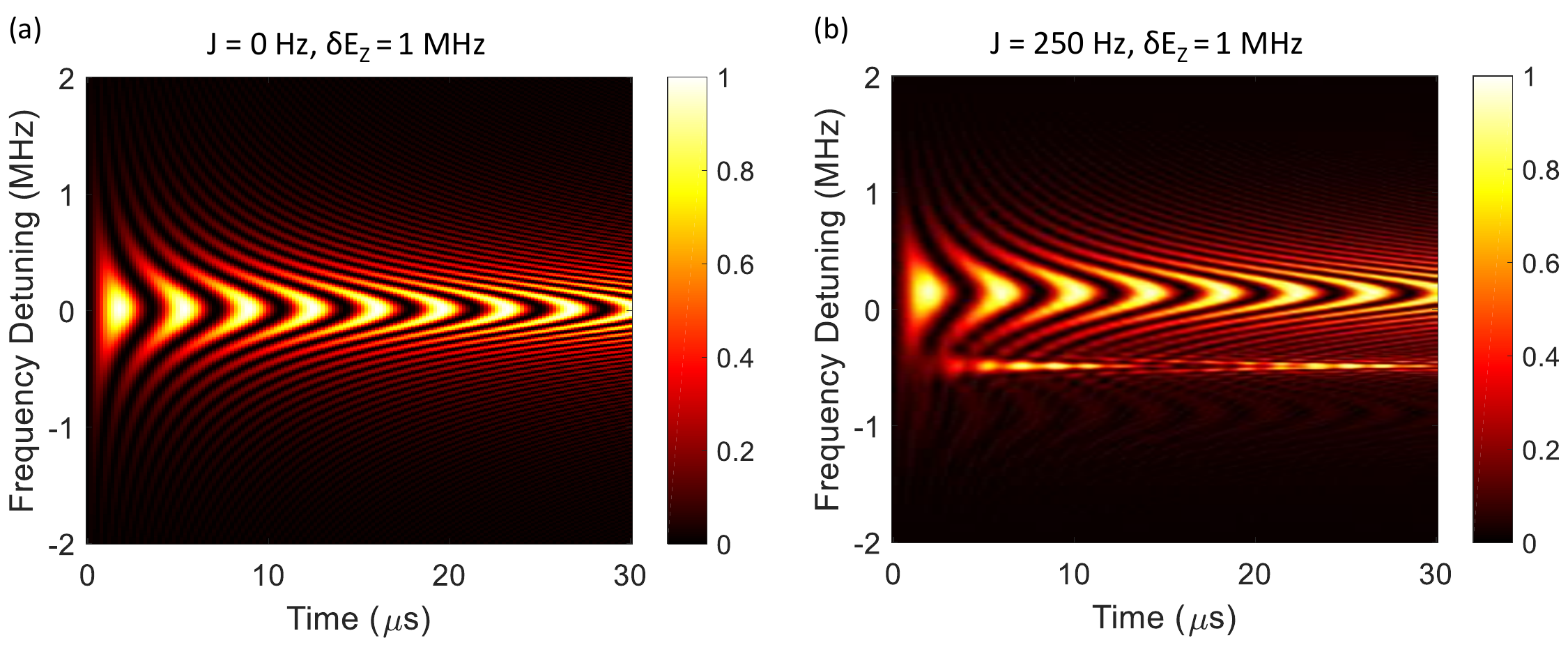}%
	\caption{Rabi simulation assuming $\delta E_Z$ = 1 MHz and (a) J = 0 Hz (b) J = 250 kHz.}	
	\label{Exchange}
\end{figure}

\begin{figure}[h]
	\includegraphics[width = \columnwidth]{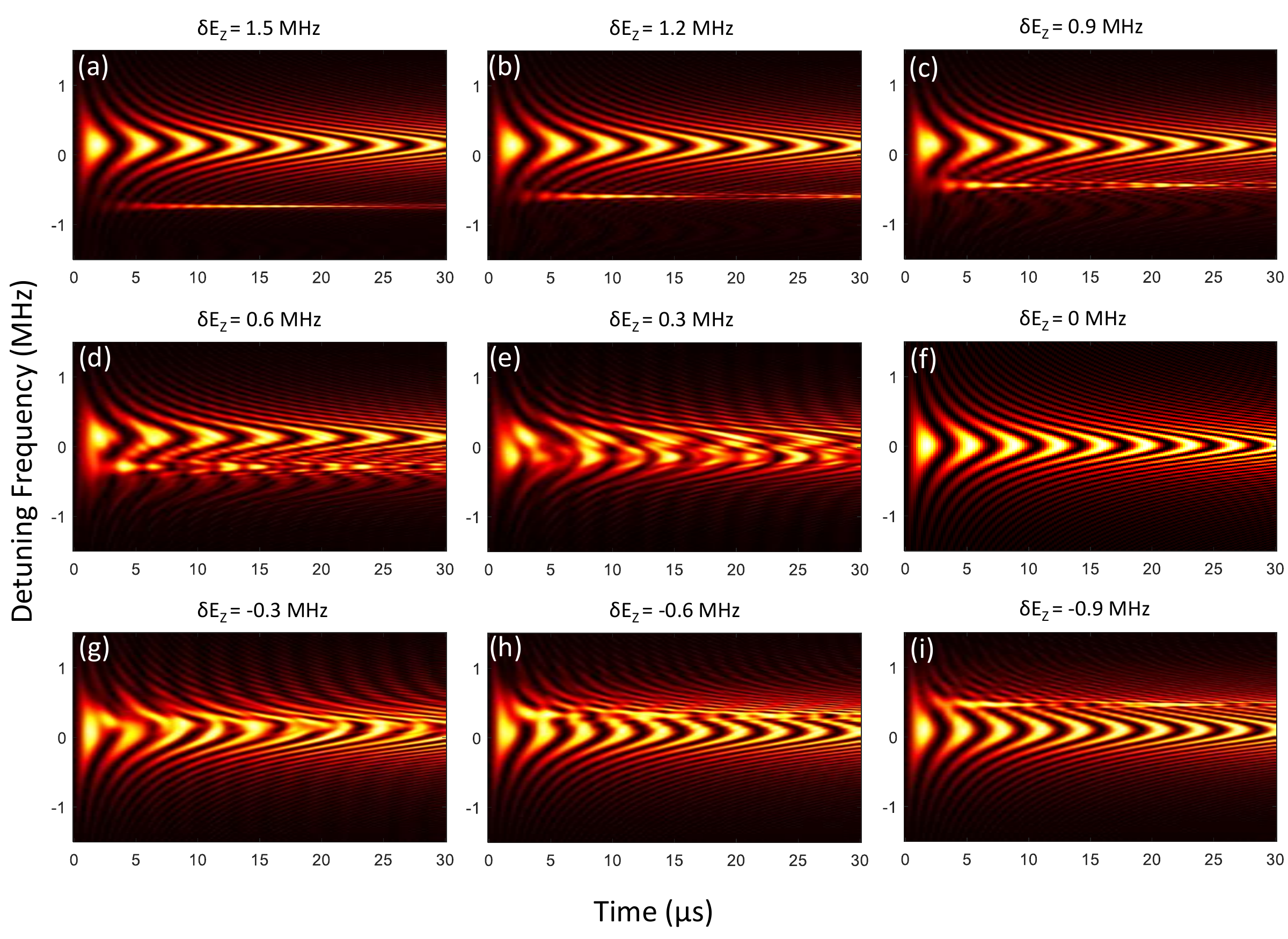}%
	\caption{Evolution of Rabi oscillation as a function of g-factor difference $\delta E_z$, with J fixed at 250 MHz .}	
	\label{gfactors}
\end{figure}

The Hamiltonian in Eq. 1 that describes two spins with Heisenberg exchange coupling is used in the Rabi simulation. By assuming a fixed $\delta E_Z$ of 1 MHz and setting J  = 0 and 250 kHz, we obtain the two Rabi maps as shown in Fig.~\ref{Exchange}.

Furthermore, by fixing J at 250 kHz, and vary $\delta E_Z$, we can observe the evolution of a narrow resonance adjacent to the main Rabi, as shown in Fig.~\ref{gfactors}. As $\delta E_Z$ decreases, the narrow resonance broadens and moves closer to the main Rabi, and then passes to the other side as $\delta E_Z$ becomes negative. \\

\section{Effective Coupling Between the $\left|\uparrow,\uparrow\right\rangle$ and $\left|\downarrow,\downarrow\right\rangle$ States}
\label{AppE}
In this section we look at the effective coupling between the $\left|\uparrow,\uparrow\right\rangle$ and $\left|\downarrow,\downarrow\right\rangle$ states. We begin with the Hamiltonian of Eq. 1. in the main text:

\begin{equation}
\left(\begin{array}{cccc}
-\Delta \omega - \frac{1}{2}\delta E_Z & \Omega & \Omega & 0 \\
\Omega & \frac{1}{2}\delta E_Z -\frac{1}{2}J & \frac{1}{2} J & \Omega  \\
\Omega & \frac{1}{2} J & -\frac{1}{2}\delta E_Z -\frac{1}{2}J & \Omega\\
0 & \Omega & \Omega & \Delta \omega + \frac{1}{2}\delta E_Z\\
\end{array}\right)
\end{equation}

We can find an effective coupling $C_{\rm eff}$ between the $\left|\uparrow,\uparrow\right\rangle$ and $\left|\downarrow,\downarrow\right\rangle$ by applying a second order perturbation approximation to Eq.~\ref{2spinH}:
\begin{equation}
\left(\begin{array}{cc}
-\Delta \omega - \frac{1}{2} \delta E_Z  & C_{\rm eff} \\
C_{\rm eff} & \Delta \omega + \frac{1}{2} \delta E_Z \\
\end{array}\right)
\label{2spinH}
\end{equation}

\begin{equation}
C_{\rm eff} = 
\frac{4 J \Omega^2 \left(J^2-2 \left({ \delta E_z}^2+2 {\delta E_z}{\Delta w}+2{\Delta w}^2\right)\right)}{(J-2{ \Delta w}) (2 { \Delta w}+J) (J-2({\delta E_z}+{\Delta w}))(2({\delta E_z}+\Delta w)+J)}
\label{Ceff}
\end{equation}

As the simultaneous rotation of the two spins occurs in the vicinity of halfway between the two qubits' resonance frequencies, Eq.~\ref{Ceff} can then be simplified by letting $\Delta \omega = -\frac{1}{2} \delta E_Z$:  
\begin{equation}
C_{\rm eff} \approx 
\frac{4J\Omega^2}{\delta E_Z^2 - J^2}
\end{equation}


\end{document}